\documentclass[runningheads]{llncs}
\usepackage{hyperref}       
\usepackage{url}            
\usepackage{booktabs}       
\usepackage{amsfonts}       
\usepackage{nicefrac}       
\usepackage{microtype}      
\usepackage{xcolor}         

\usepackage{graphicx} 
\usepackage{epstopdf} 
\usepackage{enumitem} 
\usepackage{amsmath,amssymb} 
\usepackage{bbm} 
\usepackage{graphicx}
\usepackage[linesnumbered,ruled,vlined]{algorithm2e} 

\begin{document}
%
\title{E2EAI: End-to-End Deep Learning Framework for Active Investing}

%

\author{Zikai Wei \inst{1} \and
Bo Dai\inst{2} \and
Dahua Lin\inst{1,2}} 

\authorrunning{Z. Wei et al.}
%
\institute{The Chinese University of Hong Kong \and
Shanghai AI Laboratory}

%
\maketitle              
\begin{abstract}

    Active investing aims to construct a portfolio of assets that are believed to be relatively profitable in the markets, with one popular method being to construct a portfolio via factor-based strategies. In recent years, there have been increasing efforts to apply deep learning to pursue ``deep factors'' with more active returns or promising pipelines for asset trends prediction. However, the question of how to construct an active investment portfolio via an end-to-end deep learning framework (E2E) is still open and rarely addressed in existing works. In this paper, we are the first to propose an E2E that covers almost the entire process of factor investing through factor selection, factor combination, stock selection, and portfolio construction. Extensive experiments on real stock market data demonstrate the effectiveness of our end-to-end deep leaning framework in active investing.
   
\end{abstract}

\section{Introduction}
Active investing is one of the dominant investment styles in the financial market that aims to generate excess returns compared to a market benchmark. Factor investing is one of the most common methods for active investing in a quantitative way, the core of which is using the factor as a explanatory tool for the future return. In the initial phase, human experts manually create factor databases composed of data from markets and financial reports. Then, practitioners conduct factor tests to select useful factors from the database, which later serve as candidates for building a multifactor model for stock selection. 
Finally, an active portfolio can be constructed,
which is a collection of financial investments, particularly selected stocks, that seek to provide a return greater than the return of a market benchmark, such as a major market index.
However, these steps are often optimized in isolation,
while the end-to-end active investment (E2E) is rarely addressed in existing works \cite{lin2021learning,xu2021hist,xu2021rest}.

In practice, this may lead to sub-optimal portfolios.
First, in the factor selection phase, only the factors that meet a rule of thumb are selected, e.g., the absolute value of the information coefficient (IC) should not be less than 0.02 and the absolute value of its z-score (ICIR) should not be less than 0.20. 
In this way, only those factors that have a ``sufficient'' linear correlation with the expected return are selected. However, this may result in ignoring some factors that, although useful for finding attractive stocks, are highly related to future returns and do not have a linear relationship. 
Second, the routine factor selection does not guarantee the diversity of factors. There is a risk of selecting a group of factors that have similar characteristics, leading to the problem of exposing the final portfolio to a single concentrated risk factor. 
Third, portfolio construction based on a multifactor model and the goal of maximizing weighted factor values is not a direct way to learn the optimal portfolio. A multifactor model is trained to improve the explanatory power of the inputs to the expected returns and can later be used as a tool to score stocks, and subsequent portfolio construction is based on these scores of stocks. However, the attractive stocks selected by the scores do not always result in the most profitable portfolio, as the stocks that are less attractive than the spotlights more often generate higher returns. Instead, the goal should be to achieve a higher portfolio return subject to certain constraints on stock weighting. Finally, isolated modules are later combined to construct a portfolio with a specific investment objective. However, each of these modules focuses on different learning objectives, which may distract them from achieving that objective together. In contrast, systematizing all modules in an E2E framework with a synthetic learning objective can make all modules achieve the same investment objective.

To advance the progress of deep neural networks without being hindered by the existing drawbacks of sequentially using isolated modules, in this paper we propose a novel E2E Active Investing framework (\textbf{E2EAI}) that encourages all modules, i.e., factor selection, stock selection, and portfolio construction, to fulfill the same learning objective, e.g., optimizing portfolio returns subject to weighting or risk constraints. We develop a gated attention block for both factor selection and portfolio construction. Different from the rule of thumb for factor selection in practice that selects factors satisfied the threshold of IC or ICIR, we design a gated-attention mechanism to find the factor that contributes the most in portfolio construction. Later, a deep multifactor model learns a deep factor via attention modules based on a multi-relationship stock graph that considers the intra-sector and cross-sector influence of the other stocks with different close relationships. The portfolio construction is based on the exposure of the deep factors, the market and the fundamental context of the company learned from the original inputs. The whole learning framework is trained with the same objective.

Even if an E2E framework shows promising results in active investing, portfolio managers are still interested in the financial logic and insights behind deep learning, especially how the original factors contribute to and influence the deep factor that is later used for portfolio construction. Therefore, an E2E framework should also provide an interpretation of the deep factor. To be interpretable in the deep multifactor module, we further develop a linear and directional estimator of the deep factor to identify the logic and insights behind the original factors from the perspective of linear representation. Unlike existing work \cite{wei2022factor,nakagawa2019deep,nakagawa2018deep}, we provide a directional attention mechanism that can identify the contribution of the original factor through attention while indicating the direction of the factor as positive contribution and negative contribution. The similar directional buffer can also be applied to a deep factor, which directly indicates whether the sign of its correlation with future return is positive or negative. 

To validate the effectiveness of our approach, we conduct a comprehensive study with real data containing more than 2800 constituent stocks from three broad-based indices in Chinese stock markets, where our E2E framework outperforms existing investment pipelines whose modules are separately optimized. The deep factor learned by our E2E method can also deliver a better portfolio than existing multifactor models. In summary, the contribution of this work is to 1) develop the first E2E framework for active investing using original data for portfolio construction, 2) propose a novel loss design that incorporates a directional multiplier to determine the direction of the deep factor, recover the original factor contribution, and determine the historical cross-sectional factor returns of the deep factor, 3) develop a training algorithm with a global optimizer for training the E2E framework and a local optimizer to compute the historical factor returns of the deep factor with a directional recovery.
\section{Related works}
\noindent\textit{Deep Multifactor Model Incorporating Stock Relationship.} 
The multifactor model has received considerable attention from researchers and academics for several decades to determine the exact nature of the common factors that influence risk and return in various assets and markets \cite{melas2018best}, with a large body of work showing that deep or nonlinear models perform better than linear models \cite{levin1995stock,nakagawa2018deep,nakagawa2019deep}. Moreover, cross-sectional factors can provide a better explanation for average stock returns \cite{fama2020comparing}, which inspires us to develop a deep learning architecture to learn cross-sectional factors. With the rapid development and its advantages in learning nonlinear relationships from Big Data in finance \cite{jiang2021applications}, graph neural networks have been successful in solving a variety of finance problems \cite{wang2021review}, such as stock movement prediction \cite{xu2021hist,lin2021learning,chen2019investment}, event-driven prediction \cite{xu2021rest}, and risk management \cite{lin2021deep}.\cite{velivckovic2017graph}. In our work, we design a relational neutralization block based on a gated graph attention mechanism to learn the multi-relational stock relationship from different perspectives.

\vspace{1.5mm}
\noindent\textit{Interpretation of deep factors.} 
Although deep learning methods perform well in learning deep factors in terms of the multi-factor model, the lack of transparency and limited interpretability of the deep factor make it difficult for institutional investors to apply a model that operates in a black-box manner in actual investment practice. One remedy is to use layer-wise relevance propagation (LRP) \cite{bach2015pixel}, a method that highlights which input features it uses to support its prediction \cite{montavon2019layer}, to capture the linear relationship between the input factors and the prediction \cite{nakagawa2018deep,matsunaga2019exploring}. Later, \cite{wei2022factor} provides a more direct way to learn the linear contribution of each input factor to the deep factor via an attention mechanism. However, these methods neglect the directional property of the original factor's contribution to the deep factor, \emph{i.e.}, the contribution of an original factor can be positive or negative. Without considering the different directions of the contribution, the linear approximation of the original factors in \cite{wei2022factor} may ignore the factor that contributes negatively to the deep factor, resulting in a ``biased'' approximation. In this paper, we propose a directional linear approximation of the deep factor via an attention module and a directional buffer to account for both the positive and negative contribution of the original factor to the deep factor.

\vspace{1.5mm}
\noindent\textit{Portfolio construction.} In existing work, portfolio construction is usually performed by the ad hoc methods, \emph{e.g.}, dividing stocks into ten groups with equal numbers \emph{w.r.t} their factor exposures and selecting the top or bottom group as the portfolio \cite{wei2022factor,duan2022factorvae}. In our work, we use a soft allocation strategy to simultaneously select attractive stocks and assign them an appropriate weight in the portfolio.
{Some applications use deep learning methods to solve the portfolio allocation problem \cite{uysal2021end,wang2019conservative} in an ``E2E'' way. However, they solves the problem only in the context of portfolio construction, which is one and the last module in our E2E framework. To best of our knowledge, we are the first to solve the problem of active investing in a holistic but also E2E way.}



\begin{figure}[t]
\centering
\includegraphics[width=0.85\textwidth]{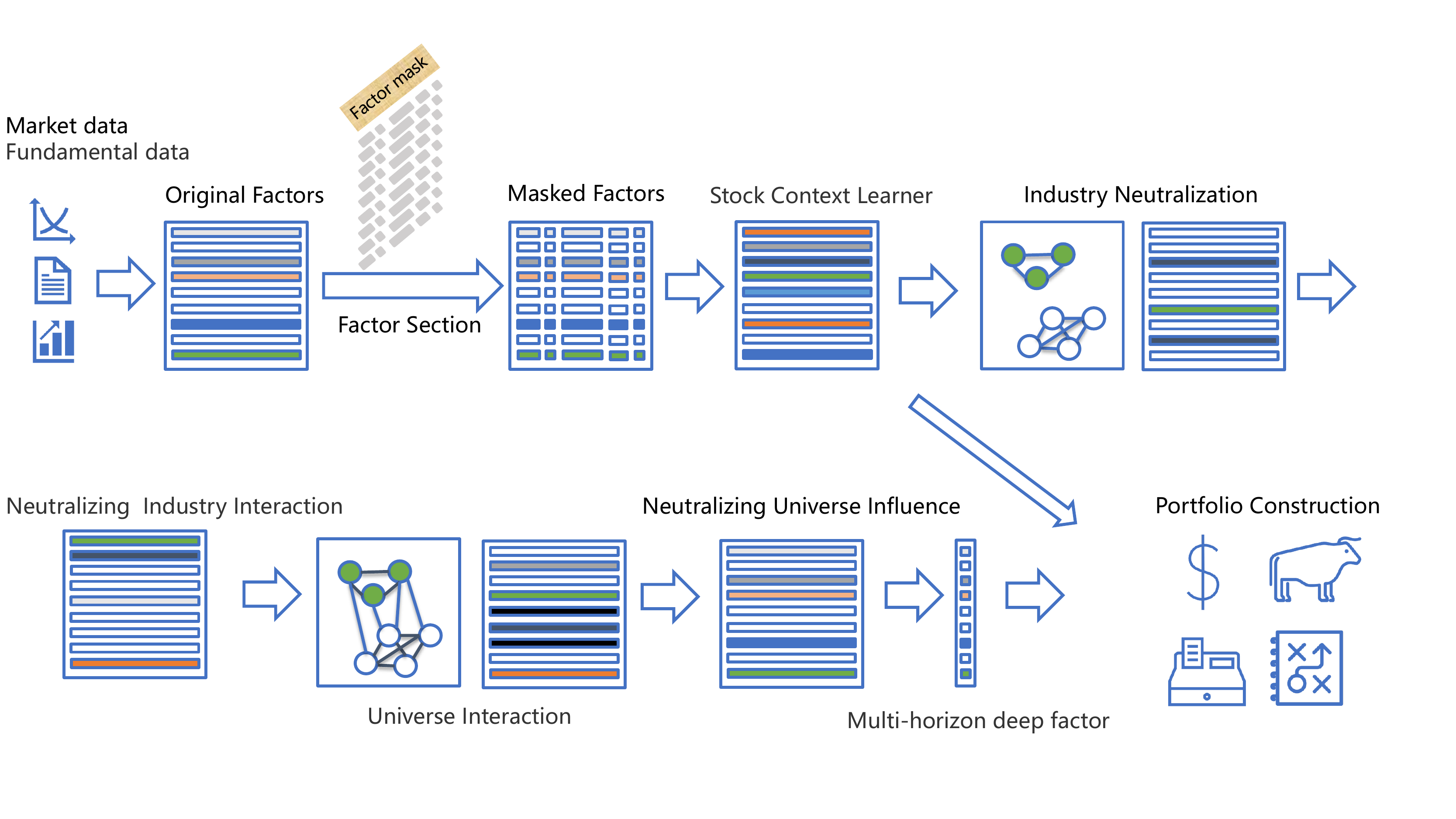}
\vspace{-0.85cm} 
\caption{The pipeline of E2E active investing framework (E2EAI).} \label{fig1_pipeline}
\vspace{-0.5cm}
\end{figure}

\vspace{-4mm}
\section{Preliminaries} \label{graph_construction}
\vspace{-2mm}
\begin{definition}
    The problem of active investing is defined as follows: Given a set of original factors $\mathcal{F}_t=\{\mathbf{f}_{it} \in \mathbb{R}^n, i =1, \dots,m\}$ with $m$ being the number of factors. Each factor is a handcrafted characteristic of $n$ stocks, with values changing over time. These factors can be engineered from market data or fundamental data (which come from financial reports).
    One should learn a stock allocation $\mathbf{w}_t = [w_{1t}, \dots, w_{nt}]^{T}$ to maximize the future portfolio return from time $t$ to $t+k$, $r_{p}^{t+k}=\mathbf{w}_t \hat{\mathbf{r}}_{t+k}$, with a set of constraints $\mathcal{C}$. $\hat{\mathbf{r}}_{t+k}=\mathbb{E}[\mathbf{r}_{t+k}|\mathcal{F}_t]$ is the expected return of each stock in the stock pool and $\mathbf{r}_{t+k} \in \mathbb{R}^n$. Some common constraints are: 1) $\sum_{j=1}^{n} w_{jt}=1$; 2) $0 \le w_{jt} \le u$, where $u$ is the upper bound of a stock allocation with $0 <u \le 1$.
\end{definition}

\begin{definition} 
The stock graph \cite{xu2021rest,wei2022factor} is defined as a directed and dynamic graph as, 
$\mathcal{G}_t = \langle \mathcal{S}_t, \mathcal{R}_t,\mathcal{M}_t \rangle$, where $\mathcal{S}_t$ denotes the set of constituents of a broad-based stock index and $\mathcal{R}_t$ is the set of relations between two stocks at time $t$. $\mathcal{M}_t$ is the set of adjacent matrices. For an adjacent matrix $\mathbf{M}_t^r \in \mathcal{M}_t$ of relation $r_t \in \mathcal{R}_t$, where $\mathbf{M}_t^r \in \mathbb{R}_t^{ | \mathcal{R}_t| \times | \mathcal{R}_t| }$, $\mathbf{M}^{r}(i,j)=1$ means that there is a relation $r$ from stock $s_j^{t}$ to stock $s_i^{t}$ and $\mathbf{M}_t^{r}(i,j)=0$ means that there is no such relation. 
\end{definition}

\section{Method}
In this work, we develop an E2E deep learning framework to find the cross-sectional factors that can consistently explain the average stock returns well, making good use of the relationships between stocks and the insights from finance. (see Fig.~\ref{fig1_pipeline}).


\vspace{-3mm}
\subsection{Factor Selection} \label{factor_selection}
The classical way in factor selection is to select factors that meet the IC or ICIR threshold using the rule of thumb. However, this can have two drawbacks: 1) It only considers the isolated and linear correlation between a single factor and the realized return. 2) It is a static selection procedure where the candidate list remains unchanged over a period of time. To mitigate the drawbacks of the existing selection practice, we develop a non-linear and dynamic factor selection module using an attention mechanism.
The factor selection module is defined as the follows
\vspace{-0.3cm} 
\begin{align}
    \mathbf{M}^{t}_{f} &= \mathbbm{1}_{\ge \gamma_f}\left(   \text{softmax} \left( \text{MLP} (\mathbf{F}^{t}_{o}) \right)  \right), \\
    \mathbf{F}^{t} &=\mathbf{M}^{t}_{f} \mathbf{F}^{t}_{o},
\end{align}
{\color{black}{where $\mathbbm{1}_{\ge \gamma_f}(\cdot)$ denotes an attention gate that forces the focus of the selection module to higher attention factors} and $\gamma_f$ is the attention lower bound for an original factor. $\mathbf{F}^{t}_{o} \in \mathbb{R}^{n \times m}$ is a matrix representing $m$ factors \emph{w.r.t} $n$ stocks.}

\vspace{-3mm}
\subsection{Deep Multifactor Model} \label{dfmf}
\noindent\textit{Stock Context Encoder.}
We define $n$ as the number of constituents inside a broad-based stock index, and $m$ as the number of types of stock context that covers fundamental, trading information and analysts' estimates. 
\begin{align}
    \mathbf{C}_t &= \text{MLP}\left(\text{BatchNorm}\left(\mathbf{F}_t\right)\right),
\end{align}
where $\mathbf{F}_t \in \mathbb{R}^{n \times m}$ is a matrix represents $m_s (m_s \le m)$ factors selected from original factors \emph{w.r.t} $n$ stocks
, and $\mathbf{C}_t \in \mathbb{R}^{n \times m_1}$ is the stock context matrix with $m_1$ hidden features extracted from the original factors. 
Batch normalization \cite{ioffe2015batch} is a deep learning counterpart to \emph{z-score normalization}, which is generally applied in data preprocessing for factor engineering \cite{wei2022factor}.

\vspace{1.5mm}
\noindent\textit{Relational Neutralization Block.} As a rule of thumb in investment practice, factor processing usually applies industry neutralization and size neutralization to remove the effect of industry and capitalization. Therefore, we design a relational neutralization block to remove the effect of different types of relationships. To encode the stock relationships, we apply Graph Attention Network (GAT) \cite{velivckovic2017graph} to predefined stock graphs and its corresponding stock context, which is a common paradigm in existing works \cite{wei2022factor,xu2021hist,wang2021review}.
The relational neutralization block (RNB) is defined as
\begin{align}
     \bar{\mathbf{C}}^{t} & = \text{RNB}(\mathbf{C}_{t}; \mathcal{G}_t) = \mathbf{C}_{t} - \text{GAT}(\mathbf{C}_{t}, \mathbf{M}_t^r),
\end{align}
where $\mathbf{C}_t \in \mathbb{R}^{n \times m_1}$ is the stock context, $\mathcal{G}_t$ is the given stock graph at time $t$ and $\mathbf{M}_t^r$ is its corresponding adjacent matrix representing the stock relationship $r$. Similar to \cite{wei2022factor}, we have two neutralized stock context corresponding to inner-industry and cross-industry relationship, $\bar{\mathbf{C}}_{I}^{t}$ and $\bar{\mathbf{C}}_{U}^{t}$, \emph{i.e.}
\begin{align}
     \bar{\mathbf{C}}^{t}_I & = \text{RNB}(\mathbf{C}_{t}; \mathcal{G}_t^I),  ~\bar{\mathbf{C}}^{t}_U = \text{RNB}(\bar{\mathbf{C}}^{t}_I; \mathcal{G}_t^U),
\end{align}
where $\mathcal{G}_I^t$ is the industry graph and $\mathcal{G}^t_U$ is the cross-industry graph. 

\vspace{1.5mm}
\noindent\textit{Learning Deep Factors for Multiple Horizons.} 
We design $K$ output heads for learning ultimate factors corresponding to multiple forward periods be aware of practitioners, \emph{i.e.} $k$-forward trading days, where $k=3, 5, 10, 15, 20$.  Formally, we learn the deep factor $\mathbf{f}_{k}^t \in \mathbb{R}^n$ that explains the future information on the $k$-forward trading day from different granularity: stock original context $\mathbf{C}_{}^{t}$, industry neutralized context $\bar{\mathbf{C}}_{I}^{t}$, and universe-neutralized context $\bar{\mathbf{C}}_{U}^{t}$:
\begin{align}
\label{eq_factor}
    \mathbf{f}_{k}^t = \text{LeakyReLU} \left(\mathbf{W}_{[k]}^T \left(\ \mathbf{C}_{}^{t} || \bar{\mathbf{C}}_{I}^{t} || \bar{\mathbf{C}}_{U}^{t} \right)\right),
\end{align}
where $\mathbf{W}_{[k]}^T$ is a single-layer feed-forward neural network for $k$-forward trading days and $||$ represents concatenation.

\vspace{-2mm}
\subsection{Portfolio Construction on Automatic Stock Selection} \label{dfmf}
\noindent\textit{Directional Buffer.} To determine the direction of the contribution of each original factor to the deep factor in the deep multifactor module, we develop a directional buffer to ``memorize'' the directions of each factor. This direction is defined by the sign of the accumulated linear correlations between the original factor and the deep factor in the training period: $\mathbf{d}_k^{[i]} = \left(d_1,\dots,d_m \right)$, with
$\mathbf{d}_k^{[i]} = \text{sgn} \left(\sum_{t\in \mathcal{T}} \text{IC}(\mathbf{f}_i^t, \mathbf{f}_k^t) \right)$, where $\text{sgn}$ denotes a sign function, $\mathbf{f}_i^t \in \mathcal{R}^n$ is the $i$-th factor in $\mathbf{F}_t$.
The details of the direction buffer is shown in Algo. \ref{algo}.

\vspace{1.5mm}
\noindent\textit{Interpretation of Deep Factor via Directional Attention Module.} We introduce directional factor attention module to investigate the positive or negative contribution of original input factors or features to the deep factor. We employ attention mechanism \cite{vaswani2017attention} to attend our deep factor to learn the token importance of the original features. The normalized attention weight through a softmax function illustrates ``how much information influx in the deep factor from an original feature''.
\begin{align}
    \mathbf{A}^{t}_{k} & =  \text{softmax} \left(\text{LeakyReLU} \left(\mathbf{W}_{a,k}^T  \mathbf{F}_t \right) \right), \qquad \bar{\mathbf{a}}^{t}_{k} = \frac{1}{n} \sum_{i \in n} \mathbf{a}^{t}_{ik},
\end{align}
where $\mathbf{W}_{a,k}^T \in \mathbb{R}^{n \times n}$ is a single-layer feed-forward neural network corresponding to the $k$-forward trading day. $\mathbf{A}^{t}_{k} \in \mathbb{R}^{n \times m}$ is the attention weight matrix for the deep factor $\mathbf{f}^{t}_{k}$ and $\mathbf{a}^{t}_{ik}$ is $i$-th column of $\mathbf{A}^{t}_{k}$. Here, $\mathbf{a}^{t}_{k}$ is its corresponding attention weight vector. We define \emph{the linear approximation of deep factor}, $\hat{\mathbf{f}}^{t}_{k} \in \mathbb{R}^{m}$, as 
\begin{align}
    \hat{\mathbf{f}}^{t}_{k} =  \mathbf{F}_t^T \left(\bar{\mathbf{a}}^{t}_{k} \circ \mathbf{d}_{k} \right),
\end{align}
where $\circ$ denotes Hadamard product, $\mathbf{d}_{k}\in \mathbb{R}^{m}$ is the directional buffer of original factors \emph{w.r.t} horizon $k$. When $\hat{\mathbf{f}}^{t}_{k}$ is very close to $\mathbf{f}^{t}_{k}$, the attention weight $\bar{\mathbf{a}}^{t}_{k}$ can interpret the portion of quantity comes from the original input factors.

\vspace{1.5mm}
\noindent\textit{Learn a Deep Portfolio with Automatic Stock Selection.} We develop a gated additive attention mechanism for automatic stock selection based on the deep factors generated by the previous module and the market embedding learned from all stock contexts. The attention score for all stocks in the universe is defined by Eq. (10) and the allocation of the selected stocks is shown in Eq. (11):
\begin{align}
    \mathbf{A}^{t}_{p,k} &= \text{softmax} \left(\text{LeakyReLU} \left(\mathbf{W}_{a,k}^T \left( \mathbf{C}_t ||(\mathbf{f}^{t}_{k} \circ \mathbf{d}_{f,k})\right)\right)\right), \\
    \mathbf{W}^{t}_{k} &= \text{softmax} \left(\mathbbm{1}_{\ge \gamma_p}(\mathbf{A}^{t}_{p,k}) \right),
\end{align}
where $\gamma_p \in (0,1]$ is the attention lower bound, $\mathbf{d}_{f,k}$ is the directional buffer of deep factor. If the attention weight of a stock does not satisfy this requirement, then it would not be considered in the portfolio. 

\begin{algorithm}[t]
\label{algo}
\scriptsize
\SetKwData{Left}{left}\SetKwData{This}{this}\SetKwData{Up}{up} \SetKwFunction{Union}{Union}\SetKwFunction{FindCompress}{FindCompress} \SetKwInOut{Input}{Input}\SetKwInOut{Output}{Output}
	\caption{Algorithm for Training E2EAI} 
	\Input{training data $\mathcal{D}^{\text{train}}$, model parameters $\Phi$, $\Psi$ }
	\Output{$\Phi^*$, $\Psi^*$}
        \BlankLine
        {Randomly initialize $\Phi$, $\Psi$}
        
        \While{not meet the stopping criteria}{

        $\mathcal{L}_{\text{train}} \leftarrow 0$, $l_f \leftarrow 0$, $b_f \leftarrow10^{-6}$, and $b_i \leftarrow10^{-6}, i=1,\dots,m$
        
		\For{$\mathcal{D}^{\text{batch}}\leftarrow \mathcal{D}^{\text{train}}$}{ 
	 	Obtain the deep factors via E2EAI: $\mathbf{f}^{\text{batch}},\mathbf{w}^{\text{batch}}\leftarrow \Phi(\mathcal{D}^{\text{batch}}) $ 

        $l_f \leftarrow 0$
        
        $d_f = \text{sgn}(b_f)$, where $b_f\leftarrow b_f + \text{sum} \left(\text{IC}(\mathbf{r}^{\text{batch}},\mathbf{f}^{\text{batch}}  )\right)$
        
        $d_i = \text{sgn}(b_i)$, where $b_i \leftarrow b_i  + \text{sum} \left(\text{IC}(\mathbf{r}^{\text{batch}},\mathbf{f}_{i}^{\text{batch}})\right), i= 1, \dots, m$
        
	 	\For{$k \leftarrow 1$ \KwTo $K$}{
           $l\leftarrow 0$
           
           \For{$i \leftarrow 1$ \KwTo $N_{\text{iter}}$}{
            Calculate the regression loss $l_i$ of cross-sectional model for deep factor $\mathbf{f}_k$, where $\psi_k$ is its cross-sectional coefficient. 
            \BlankLine
            $l\gets l-  l_i \cdot d_f$
            
            Update $\psi_t$ via the cross sectional optimizer $\gamma_1$ by minimizing $l$
           }
           $l_f \leftarrow l_f - mean(\psi_t \circ d_f)$
        }
        Calculate the loss for factor stability $l_s$ and the loss for attention estimate $l_e$, where $l_e$ is calculated based on the directional buffer: $d_i, i=1,\dots, m$. Then, update the train loss:
        
        $\mathcal{L}_{\text{train}} \leftarrow \mathcal{L}_{\text{train}} + l_p + \lambda_e l_e + \lambda_s l_s + \lambda_f l_f$
        
       Update $\Psi$ via the global optimizer $\gamma_2$.
 	   }
 	 } 
\end{algorithm}

\vspace{-3mm}
\subsection{Loss Design}
Now we ponder the design of loss function to fulfill the learning objectives of our E2E active investing. Specifically, the general loss function is 
\begin{align} 
\mathcal{L} = \mathcal{L}_p + \lambda_s \mathcal{L}_s + \lambda_f \mathcal{L}_f + \lambda_e \mathcal{L}_e,
\end{align}
where $\mathcal{L}_p$, $\mathcal{L}_s$, $\mathcal{L}_f$, and $\mathcal{L}_e$ are the loss terms associated with the portfolio, stability, factor returns, and attention estimates, respectively. $\lambda$'s are balancing coefficients.
In what follows, we denote $\mathcal{T}$ as the set of trading days in the training period and $\mathcal{K}$ as the set of future horizons.


\vspace{1.5mm}
\noindent\textit{Loss Related to Portfolio.} A fundamental objective in portfolio construction is to maximize expected return without violating the weighting constraints on equity allocation. Based on this objective, we express the loss function of our portfolio in two terms, \emph{i.e.}, loss of portfolio $\mathcal{L}_{p}$ return $\mathcal{L}_{ret}$ and the loss of weight $\mathcal{L}_{up}$ penalizing the allocation exceeding its upper bound $\theta$, \emph{e.g.}, $\theta = 0.10$, where $\mathcal{L}_{ret}= -\frac{1}{|\mathcal{T}||\mathcal{K}|}\sum_{ k \in \mathcal{K} } \sum_{ t \in \mathcal{T} } \mathbf{W}^{t}_{k} \mathbf{R}^{t}_{k}$ and $\mathcal{L}_{up}=\frac{1}{|\mathcal{T}||\mathcal{K}|}\sum_{ k \in \mathcal{K} } \sum_{ t \in \mathcal{T} }(\mathbf{W}^{t}_{k} - \theta)$. In short, $\mathcal{L}_{p} = \mathcal{L}_{ret} + \mathcal{L}_{up}$.


\vspace{1.5mm}
\noindent\textit{Maximizing Factor Stability.} 
The information ratio of information coefficient (ICIR) is a widely used metric for assessing the factor stability, and ICIR is defined as the mean of IC divided by its standard deviation \cite{lin2021learning,wei2022factor}.  
We use $c_k$ to denote the ICIR of the deep factor $\mathbf{f}_k^t$, where $c_k$ measures the stability of the predictive power of the deep factor $\mathbf{f}_k^t$ for the future return over the next $k$ trading days $\textbf{r}_{t+k}$. 
Thus, since ICIR is positively correlated to stability, the stability loss is defined as $\mathcal{L}_{e}=\frac{1}{|\mathcal{K}||\mathcal{T}| } \sum_{ k \in \mathcal{K} } \sum_{ t \in \mathcal{T} } d_k {c}^{t}_{k}$, where $d_k$ is the directional indicator with value of 1 or -1. 

\vspace{1.5mm}
\noindent\textit{Maximizing Factor Return.} 
Factor returns are the cross-sectional regression coefficients \cite{lin2021deep,wei2022factor} that indicate the return attributable to a particular common factor:
$\hat{\mathbf{r}}_{t+k} = {b^{t}_{k}} \mathbf{f}^{t}_{k} $,
where ${b}^{t} \in \mathbb{R}$ is the factor return at time $t$. Since a deep factor is expected to have a higher cumulative factor return, the loss for factor return is defined as $\mathcal{L}_f = \frac{1}{|\mathcal{K}||\mathcal{T}| } \sum_{ k \in \mathcal{K} } \sum_{ t \in \mathcal{T} }\psi^{t}_{k} d_f^k$. 

\vspace{1.5mm}
\noindent\textit{Improving ``Attention Estimate''.} We use $L_2$-norm to evaluate the deviation between the deep factor $\mathbf{f}^{t}_{k}$ and its corresponding attention estimate $ \hat{\mathbf{f}}^{t}_{k}$, which can be calculated as ${e}^{t}_{k} = {||\mathbf{f}^{t}_{k}- \hat{\mathbf{f}}^{t}_{k}||}_2$. Therefore, the attention estimate loss is defined as $\mathcal{L}_e = \frac{1}{|\mathcal{K}||\mathcal{T}| } \sum_{ k \in \mathcal{K} } \sum_{ t \in \mathcal{T} } {e}^{t}_{k}$.

\vspace{-2mm}
\subsection{Optimization}
To improve the performance of the E2E framework,
besides the global optimizer we further propose a specialized optimizer, named cross-sectional optimizer, for learning the directional cross-section coefficient. 

\vspace{1.5mm}
\noindent\textit{Cross-sectional optimizer.} 
The cross-sectional optimizer $\gamma_1$ is a local optimizer for learning the factor return of the deep factor, which is defined as the cross-sectional regression coefficient (factor return) at each time $t \in \mathcal{T}$. A linear regression model is fitted by minimizing the mean squared errors of the predicted returns and the target returns. Since a deep factor has a positive or negative IC, the factor's return has the same direction as its IC. Therefore, we first determine the direction of the deep factor and then estimate its factor using the cross-sectional optimizer. This cross-sectional optimizer is independent of the global optimizer we use to train our E2E framework.

\vspace{1.5mm}
\noindent\textit{Global optimizer.} The global optimizer $\gamma_2$ is designed to update the parameters of our deep E2E framework. It is set to simultaneously optimize the parameters of the factor selection, multifactor model, directed contribution estimator, and portfolio construction module.


\vspace{-3mm}
\section{Experiment Results}

\vskip -0.3cm
\begin{figure}[!htb]
   \centering
\begin{tabular}{ccc}
\includegraphics[width=0.31\textwidth]{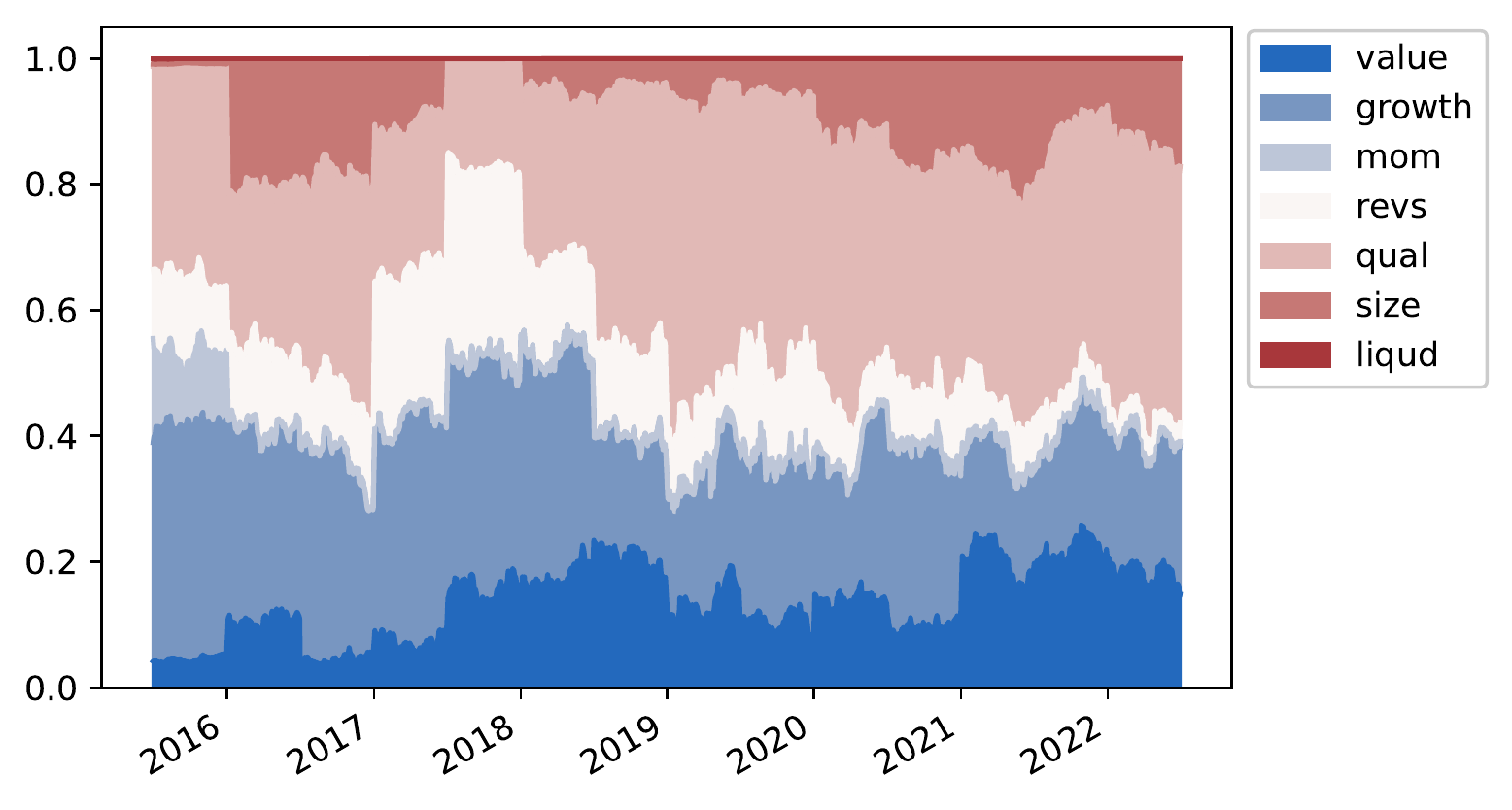}
&
\includegraphics[width=0.31\textwidth]{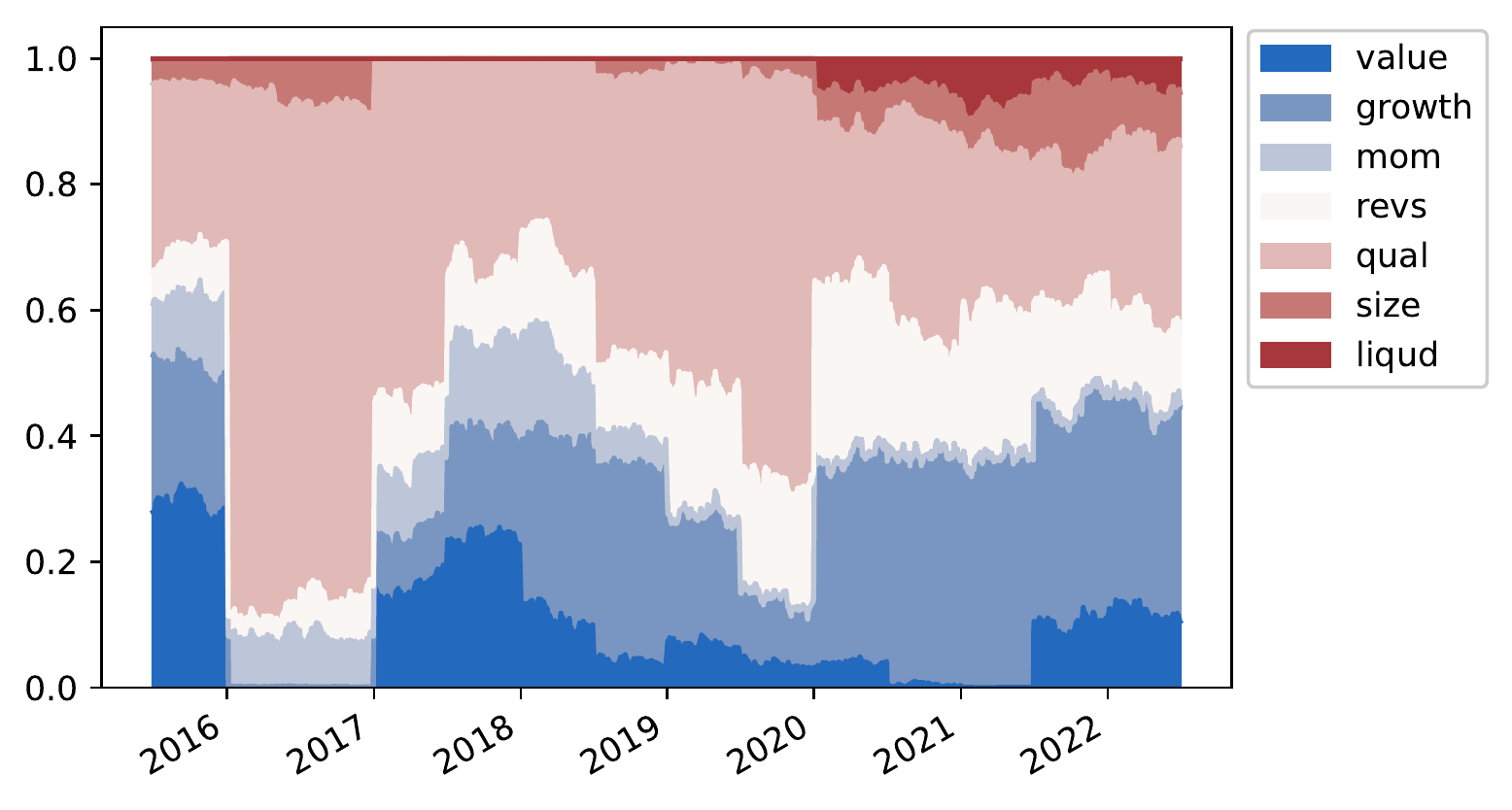}&
\includegraphics[width=0.31\textwidth]{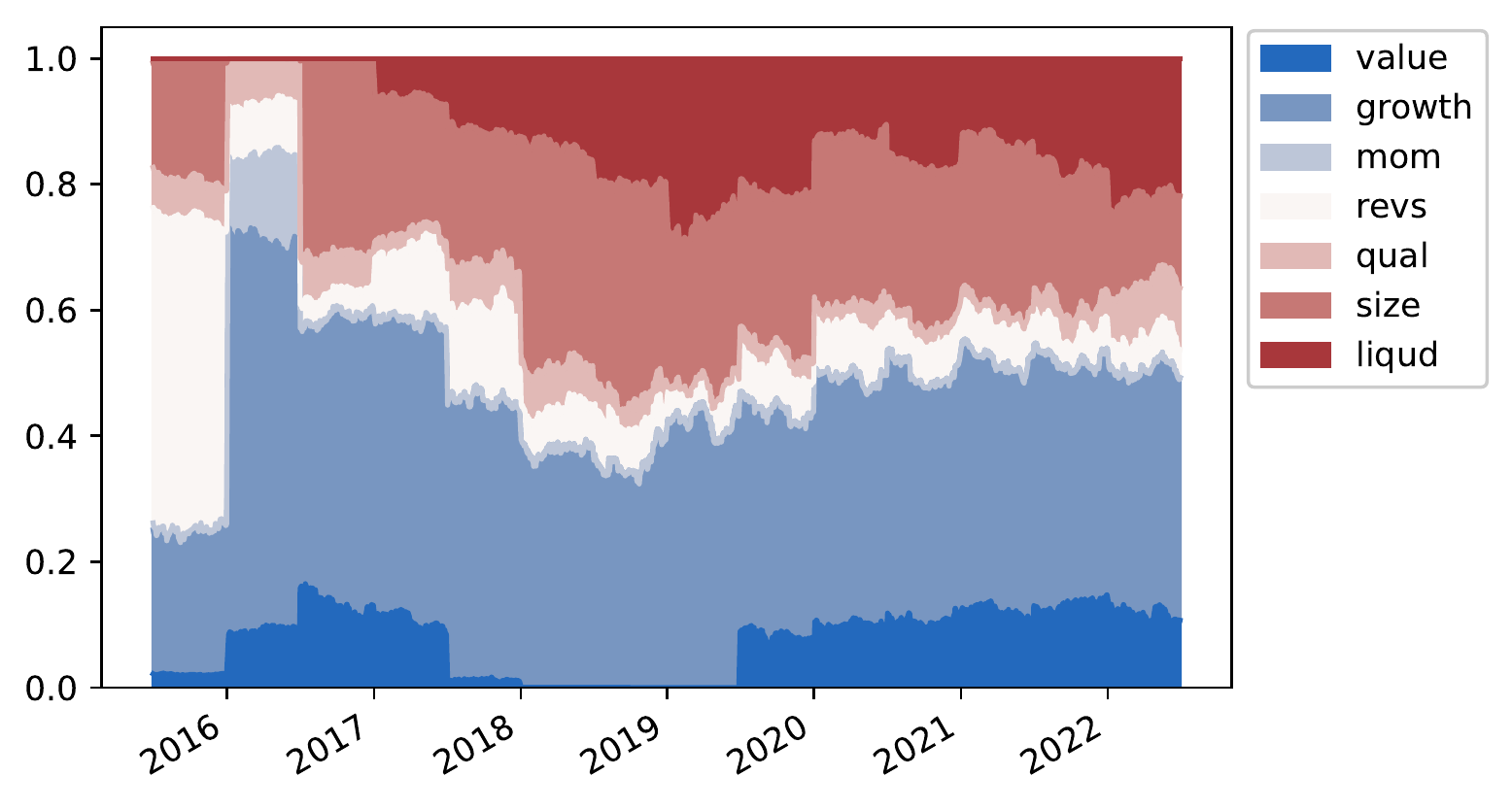}\\
\includegraphics[width=0.31\textwidth]{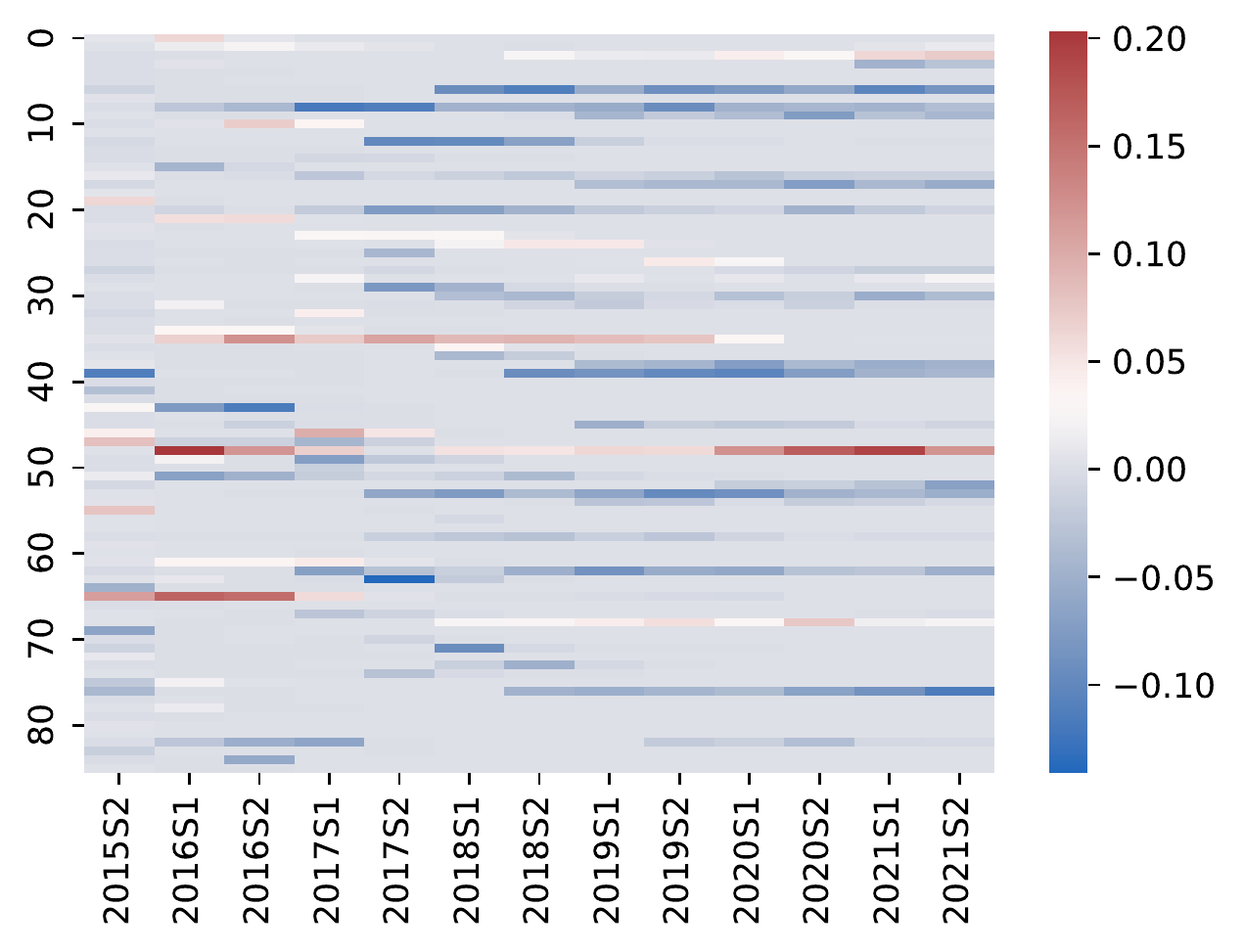}&
\includegraphics[width=0.31\textwidth]{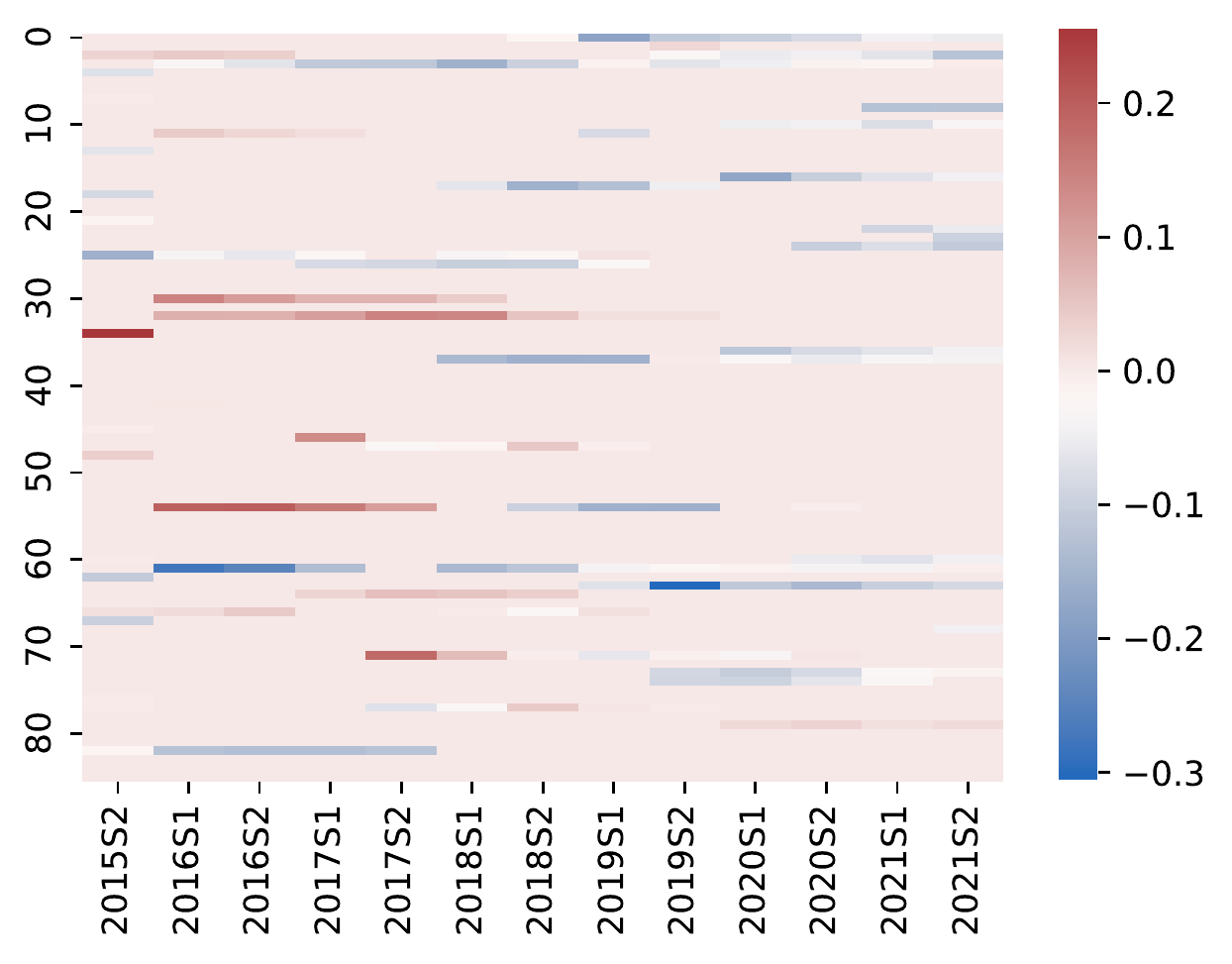}&
\includegraphics[width=0.31\textwidth]{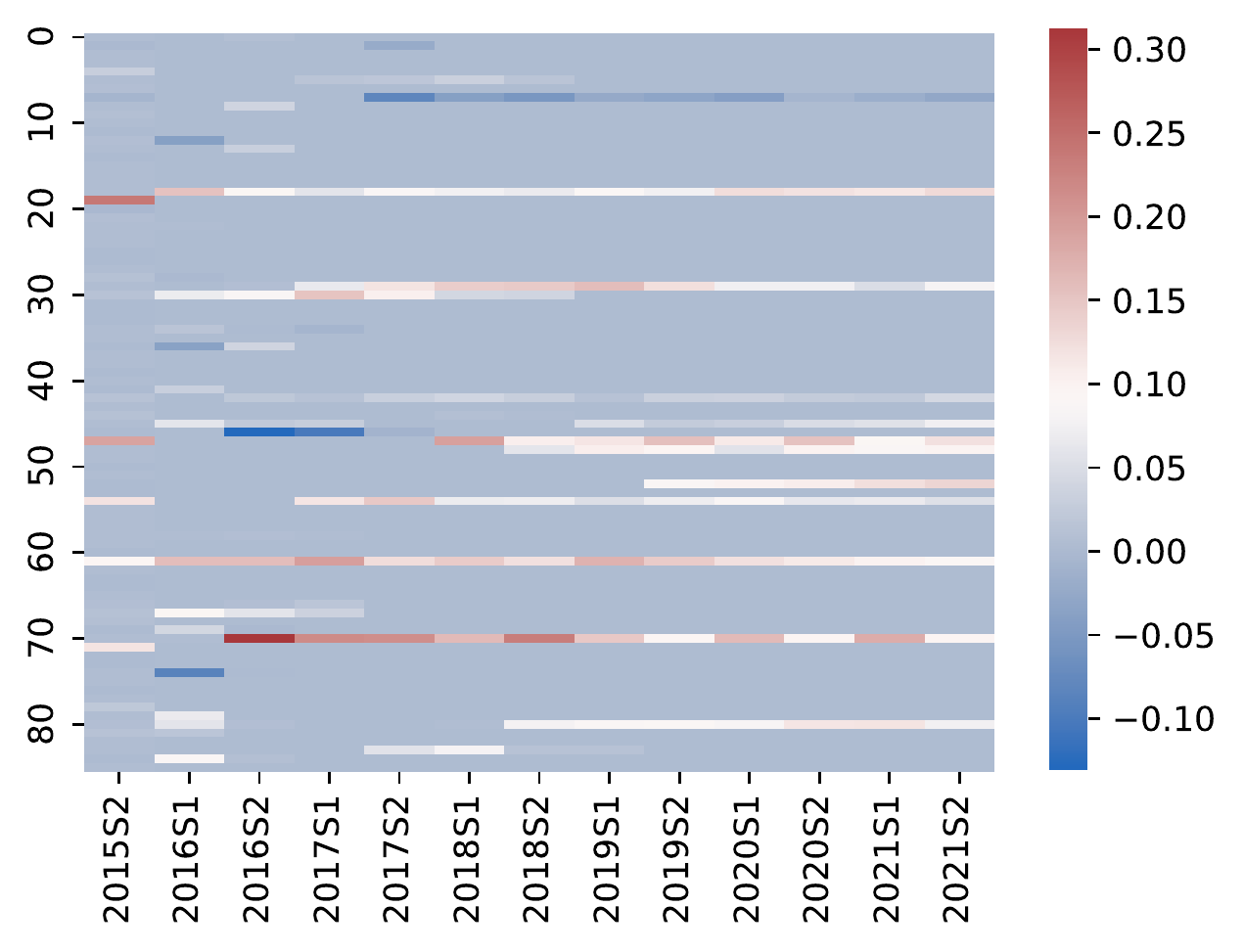}\\
CSI300 & CSI500 & CSI1000\\
\end{tabular}
\vskip -0.2cm
    \caption{The heat map shows the dynamic contribution of the original factors to the deep factors: 1) the first row shows the dynamic attention allocation to the original factor groups; 2) the second row shows the average attention weights for each style factor on a semi-annual basis.}
    \label{fig:heat_map1} 
\end{figure}
\vspace{-11mm}

\begin{figure}
   \centering
\begin{tabular}{ccc}
\includegraphics[width=0.31\textwidth]{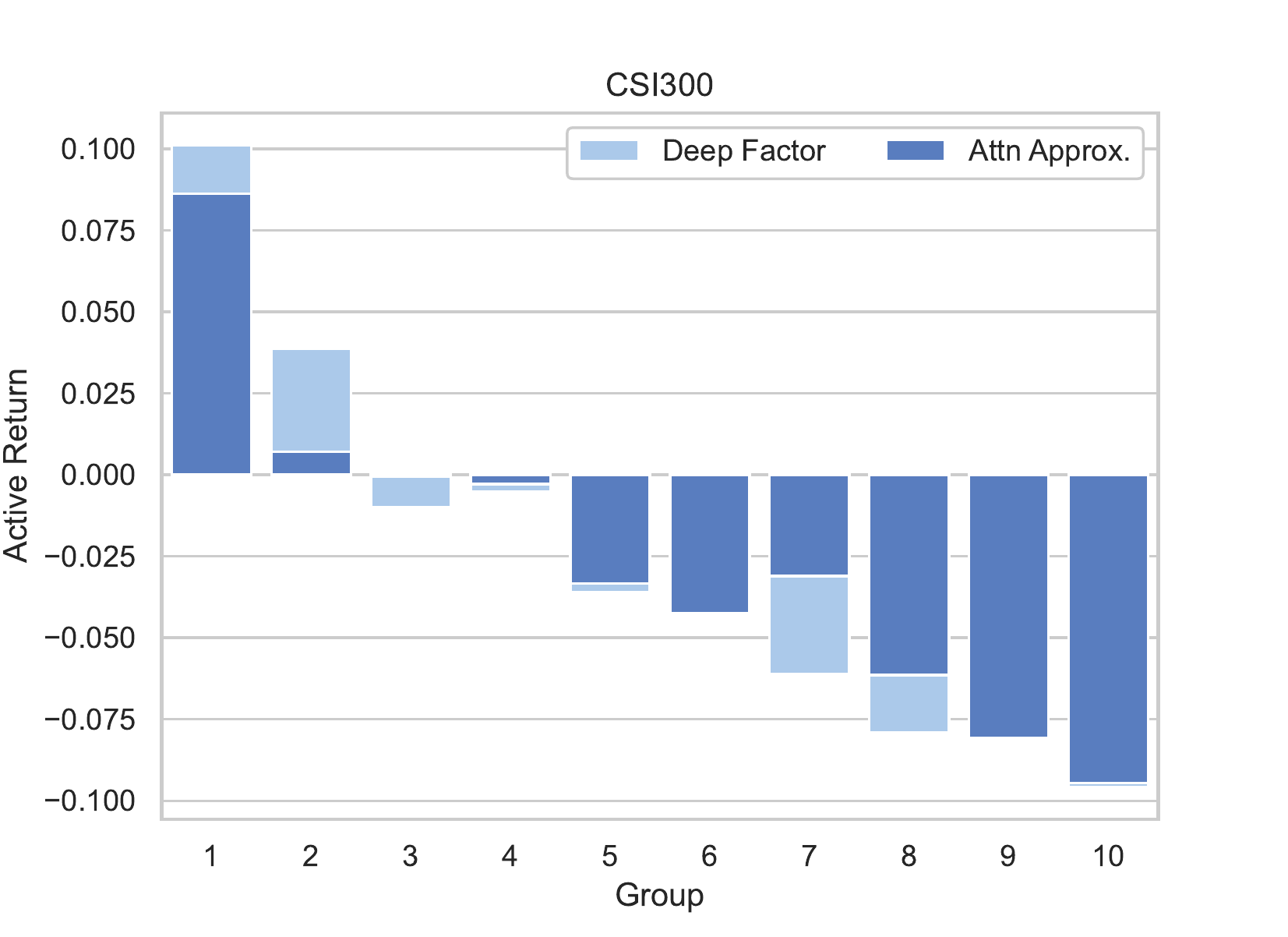}
&
\includegraphics[width=0.31\textwidth]{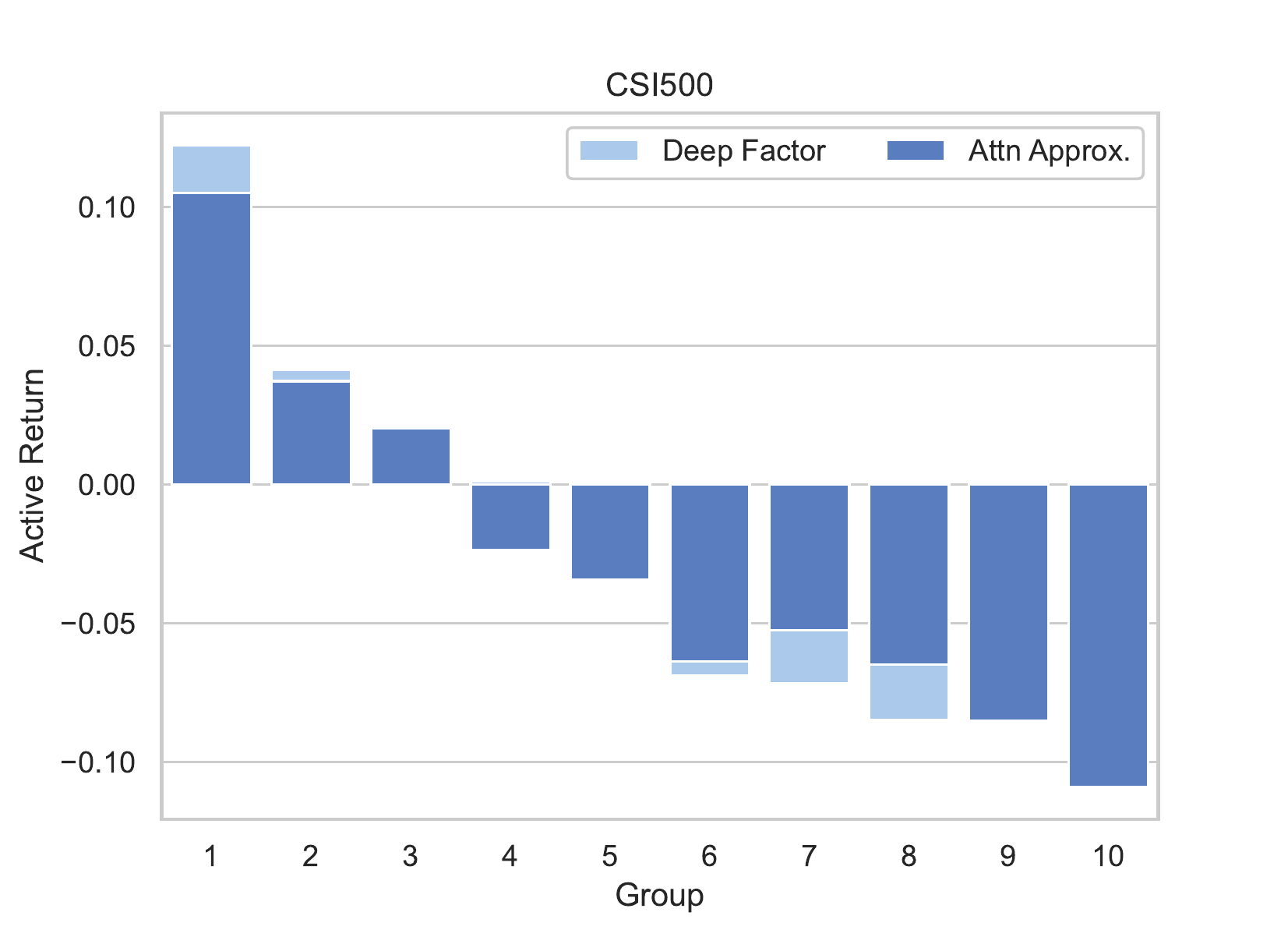}&
\includegraphics[width=0.31\textwidth]{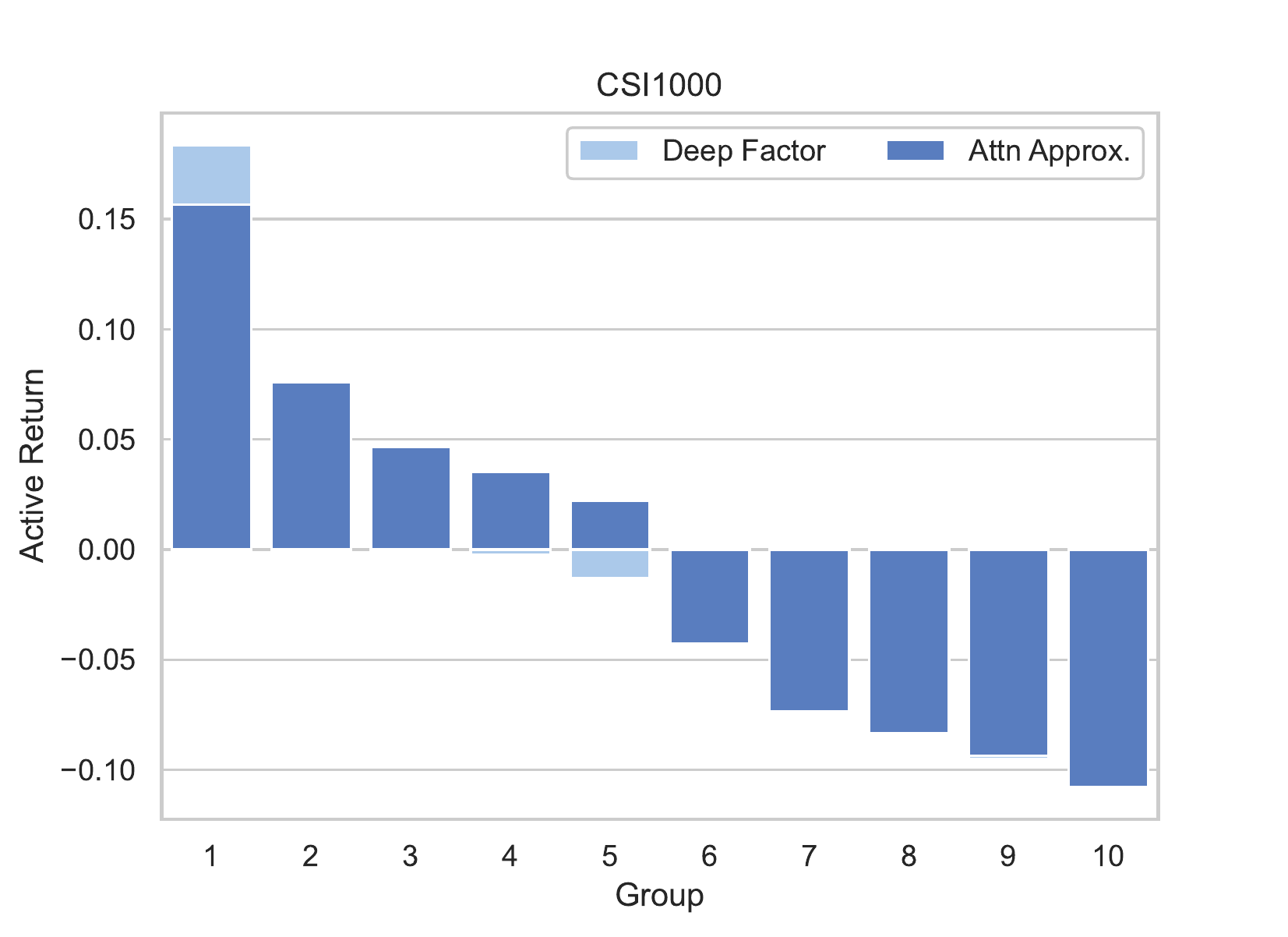}\\
\end{tabular}
\vskip -0.4cm
    \caption{The monotonicity analysis of the deep factor and its attention approximation (AA) on different stock universes. 1) The average returns increase monotonically in stratified groups based on factor exposure, showing good stability of their predictive power; 2) the AA of deep factor can linearly explain most cases in CSI300 and CSI500 from the perspective of portfolio performance in different groups.}
    \label{fig:heat_map2} 
\end{figure}

\begin{table}
  \caption{The comparison of portfolio performances based on both ad-hoc and automatic construction over the period 2015 to mid-2022 (\textbf{best}/\underline{2nd best}). 
  }
   \vskip -0.3cm
  \label{eval-metrics-method1}
  \centering
  \begin{tabular}{lccccccccccccccc}
    \toprule
    &\multicolumn{5}{c}{CSI 1000}  &\multicolumn{5}{c}{CSI 500}   &\multicolumn{5}{c}{CSI 300}                 \\
    \cmidrule(r){2-6} \cmidrule(r){7-11} \cmidrule(r){12-16}
    Method     & $\alpha$$\uparrow$ &IR$\uparrow$ & MD$\downarrow$  & TT &$\bar{n}$ & $\alpha$$\uparrow$ &IR$\uparrow$ & MD$\downarrow$ & TT &$\bar{n}$ & $\alpha$$\uparrow$ &IR$\uparrow$ & MD$\downarrow$& TT &$\bar{n}$\\
    \midrule
    Linear   & 7.79  &1.35 &0.08 & 0.76 &100   &5.62  &0.77 & \textbf{0.04} &0.89 &50 &1.73  &0.32 &0.08 &0.89 &30\\
    EW       & 14.65  &1.92& \textbf{0.06} & 0.71 &100     &8.91 &1.86 & \underline{0.06} &0.97 &50 &7.57 &1.20 &\underline{0.06}  &0.71&30\\
    MLP      & 15.16 &\underline{3.53} & \underline{0.07} & 0.88 &100     &4.48  &0.67 &0.05 &0.78 &50 &3.56 &0.65 &0.15  &0.88&30\\
    MGAT     & 9.24  &1.63 & 0.15 & 1.05 &100 &4.96 &0.62 &0.12 &0.83  &50 &3.17  &0.56 &0.09 &1.53 &30\\
    DMFM     & \underline{16.36} &3.46 & 0.08  & 0.92 &100 &9.03  &1.89 &0.07 &1.01 &50 &7.85  &1.09 &\textbf{0.06} &0.96 &30\\
    $\text{E2E}_l$ & 15.14  & 2.92 & 0.07 & 0.94 &100 &  \underline{12.00} & \underline{2.18} & 0.07  & 1.11 &50 & \underline{9.60}  &\underline{1.46} & 0.09 & 0.71 &30\\
    $\text{E2E}_d$ & \textbf{18.10} & \textbf{4.35} & 0.08  & 0.92 &100 & \textbf{13.37}  & \textbf{2.45} & 0.08& 1.01 &50 & \textbf{13.40}&\textbf{2.02} & 0.07  & 0.97 &30\\
    \midrule
    S-Best  & 7.10  &  0.42 &0.11 &  0.81  &319 &0.74  &0.90 & 0.01 &0.24 &433 &-0.06 &-0.58 &\underline{0.01}  &0.16 &272\\
    S-Avg   & \underline{11.10}  & \underline{0.67} &\underline{0.11}  & 0.82 &321 &\underline{0.88} &\underline{1.16} & \underline{0.01}  &0.22 &455 &0.01  &-0.19 &\textbf{0.01} &0.13 &289\\
    S-T20 & -0.04 &-0.08& 0.32 &1.75 &32 &-9.44 &-0.81 & 0.22  &1.03 &25 &\textbf{11.22}  &\underline{0.69} &0.11 &0.40 &50\\
    $\text{E2E}_{auto}$ & \textbf{20.12} & \textbf{1.37} & \textbf{0.09} & 0.96 &95 & \textbf{8.88} & \textbf{2.47} & 0.04  & 0.76 &129 & \underline{7.40} &\textbf{1.56} & 0.04 & 0.60  &103\\
    \bottomrule
  \end{tabular}
  \vskip -0.3cm
\end{table}

\noindent\textit{Stock Pool.} To have a fair comparison to the existing deep multifactor model, we use the same stock pool with the same period used in \cite{wei2022factor}. The stock pool consist of the most representative stocks of China stock markets, \emph{i.e.}, the constituents of three major China Stock Indices (CSIs): CSI 300, CSI 500 and CSI 1000, which represent more than 80\% of total market capitalization of the entire stock market in China \cite{wei2022factor}. These three indices covers 1800 stocks largest and most liquid stocks in the market at the same time, while they cover more than 2800 stocks in the period from 2010 to mid-2022. 

\vspace{1.5mm}
\noindent\textit{Dataset Construction.} The factors consist of fundamental factors and technical factors based on various data sources. The fundamental and market data come from publicly available data from \url{tushare.pro}, which can also be obtained from Refinitiv or WIND institutional investor databases \cite{wei2022factor}. We create our dataset on a daily basis and divide the original factors into seven groups: Value, Growth, Momentum, Quality, Size, Liquidity \cite{wei2022factor,nakagawa2019deep,melas2018best}. The daily stock return is calculated by $\mathbf{r}_{t+k}=\frac{\mathbf{p}_{t+k}-\mathbf{p}_{t+1}}{\mathbf{p}_{t+1}}$, where $t$ is the forecast day and $\mathbf{p}_{t+k}$ denotes the volume-weighted average price of all stocks on the next $k$ day after the forecast day. To avoid look-ahead bias, we create a point-in-time dataset \cite{wei2022factor} and delete the intersection of the validation and test sets from the validation set. For the stock graphs, the industry graph is created based on the CITIC Securities Industry Classification Standard \cite{wei2022factor} and the universal graph is created based on the list of constituent members issued by the China Securities Index Company. In creating the dataset, we use the time- series cross-validation technique, which divides the entire dataset into 14 groups in chronological order \cite{wei2022factor}.

\vspace{1.5mm}
\noindent\textit{Baselines and Model Implementation.} We compare our proposed method with the following baselines: 1) \textbf{Linear}\cite{fama2020comparing} is a linear multifactor model. 2) \textbf{EW} \cite{melas2018best} is an equally weighted model that has nothing to do with data mining or in-sample optimization. In practice, any deviations from equal weighting are determined by economic rationality \cite{melas2018best}. 3) \textbf{MLP} is a nonlinear model that learns factors from the stock context based on neural networks and consists of a context encoder and a feature decoder. 4) \textbf{MGAT} \cite{wei2022factor} is a model based on a context encoder, a GAT, which learns the interactions between stocks in a universe graph, and a decoder, which learns a deep factor from the stock context and the universal relationships of the stocks. 5) \textbf{DMFM} \cite{wei2022factor} is a deep multifactor model. {\color{black}{For comparison of E2E and step-by-step methods, we provide three additional step-by-step baselines with different criteria for factor selection: 
1) \textbf{S-Best}, which selects the factors with the best performance at different prediction horizons; 2) \textbf{S-Avg}, which selects the factors that can perform well on average at different prediction horizons. 3) \textbf{S-T20} selects only the factors that perform well in predicting one-month returns.}
}

{\color{black}{ 
\vspace{1.0mm}
\noindent\textit{The Performance based on Ad-Hoc and Automatic Portfolio Construction.} We use three widely used metrics for portfolio construction: excess return ($\alpha$), information ratio (IR), and maximum active drawdown (MD), as well as two additional descriptive metrics such as buy and sell turnover (TT) and average number of stocks held. $\uparrow$ and $\downarrow$ denote the metrics ''the higher the better'' and ''the lower the better'' respectively. 1) The \textit{ad-hoc method} is a popular method used to test the ability of a multifactor model to select stocks and to show the effectiveness of our multifactor model. It involves dividing stocks into $K$ groups \emph{w.r.t} of factor exposures and constructing an equally weighted portfolio based on the group with the largest factor exposure (when the factor is positively linearly correlated with expected return and vice versa). Table \ref{eval-metrics-method1} shows the comparison of the different methods based on a stratification of 10 groups, where $\text{E2E}_d$ and $\text{E2E}_l$ denote the portfolio constructed based on a deep factor and its linear approximation (AA) via attention estimate. 
2) In Table \ref{eval-metrics-method1}, we compare the stepwise with the end-to-end portfolio construction, and our E2E method outperforms the other stepwise baselines. Overall, our E2EAI can outperform different baselines under different portfolio construction paradigms \emph{w.r.t} metrics such as  $\alpha$ and IR.
}}

\vspace{1.5mm}
{\color{black}{ 
\noindent\textit{Interpreting Deep Factor via Directional Contribution.} 
1) Learning the directional contribution can reveal the dynamic allocation of the original factors.
Fig. \ref{fig:heat_map1} shows the group contribution and the directional contribution of each factor over the test period. An interesting result is related to liquidity: liquidity hardly contributes in the universe of CSI300, while in CSI500 it contributes a throw in the second tier of the most liquid stocks with the largest market capitalization. However, liquidity contributes more in CSI1000 as it consists of 1000 small cap companies that are less liquid than CSI 800 (CSI300 and CSI500). This shows that liquidity is more important in small caps, which makes sense in practice.
2) The portfolio $\text{E2E}_l$ constructed by attention approximation over attention estimate is monotonically similar to $\text{E2E}_d$, see Fig. \ref{fig:heat_map2}. }}


\section{Conclusion}
\vskip -0.3cm
In this paper, we are the first to propose an E2E framework for active investing that outperforms existing methods in three major stock universes in the Chinese stock markets. Using a directional buffer, we can identify both the positive and negative contribution of each factor and also determine the dynamic allocation of each orignal factor. In the future, we plan to apply our method to the entire market with diversified client-oriented investment objectives.
\bibliographystyle{splncs04}
\bibliography{pakdd_submission_12}
\end{document}